\documentclass[a4paper,fleqn,usenatbib]{mnras}

\usepackage{mathptmx}

\usepackage[T1]{fontenc}
\usepackage{ae,aecompl}

\usepackage{graphicx}	
\usepackage{amsmath}	
\usepackage{amssymb}	

\title[Are Pebble Pile Planetesimals Doomed?]{Are Pebble Pile Planetesimals Doomed?}

\author[T. Demirci et al.]{
Tunahan Demirci$^{1}$\thanks{E-mail: tunahan.demirci@uni-due.de (TD)},
Maximilian Kruss$^{1}$,
Jens Teiser$^{1}$,
Tabea Bogdan$^{1}$,\newauthor
Felix Jungmann$^{1}$,
Niclas Schneider$^{1}$
and Gerhard Wurm$^{1}$
\\
$^{1}$	University of Duisburg-Essen, Faculty of Physics, Lotharstr. 1, 47057 Duisburg, Germany
}

\date{Accepted 2019 January 7. Received 2018 December 18; in original form 2018 August 14}

\pubyear{2019}

\begin{document}
\label{firstpage}
\pagerange{\pageref{firstpage}--\pageref{lastpage}}
\maketitle

\begin{abstract}
In parabolic flight experiments we studied the wind induced erosion of granular beds composed of spherical glass beads at low gravity and low ambient pressure. Varying g-levels were set by centrifugal forces.
Expanding existing parameter sets to a pressure range between $p=300-1200\,$Pa and to g-levels of $g=1.1-2.2\,\rm m\,s^{-2}$ erosion thresholds are still consistent with the existing model for wind erosion on planetary surfaces by \cite{Shao2000}. These parameters were the lowest values that could technically be reached by the experiment. The experiments decrease the necessary range of extrapolation of erosion thresholds from verified to currently still unknown values at the conditions of planetesimals in protoplanetary discs. We apply our results to the stability of planetesimals. In inner regions of protoplanetary discs, pebble pile planetesimals below a certain size are not stable but will be disassembled by a head wind.
\end{abstract}




\section{Introduction}
\label{sec:introduction}

Explaining the formation of planets is still a constant tuning and fine-tuning of different processes along the main route. Terrestrial planets are thought to be built bottom-up, starting with small dust grains of micrometre size \citep{Blum2008}.
These grains initially collide gently, stick together and grow to larger aggregates \citep{Blum2008,Guettler2010}. Once they reach millimetre size at collision velocities still well below $1\,\rm m\,s^{-1}$ they do not fragment yet but neither grow. They get compacted and essentially only bounce off elastically \citep{Weidling2009,Zsom2010,Jankowski2012}. Being a first natural obstacle to further growth this is called bouncing barrier \citep{Zsom2010}. A couple of laboratory experiments have confirmed that this might be a rather stable size limit \citep{Kelling2014,Kruss2016,Kruss2017,Demirci2017}.

Under specific assumptions of nanometre water ice grains \citet{Okuzumi2012} proposed that fluffy icy planetesimals can grow directly.  Also, seeding the reservoir of mm-aggregates with a few larger bodies allows these larger objects to grow in collisions with the small ones at higher speed \citep{Wurm2005,Teiser2009,Windmark2012,Windmark2012VD,Meisner2013}.
However, a continuous growth via collisional sticking to kilometre sized planetesimals seems difficult in general and  reservoirs of small, compact dust granules of mm-size seem to be prevailing.
Some tuning of the growth might be possible by considering specific materials e.g. water ice \citep{Gundlach2011,Aumatell2014,Gaertner2017}, high temperatures \citep{deBeule2017,Demirci2017} or electric charge \citep{Jungmann2018} to shift the bouncing limit maybe to centimetre size particles. It should be noted that this remains one of the challenges.

It is challenging in the sense that the alternative scenarios for creating planetesimals require a minimum particle size to evolve them further.
One promising mechanism is the streaming instability \citep{Youdin2005,Johansen2007}. In numerical simulations it turns out that particles of certain size can be concentrated in protoplanetary discs due to the interaction between solid particles with the surrounding gas. In absolute numbers centimetre to metre particles are needed in the inner disc regions while far out millimetre size might be sufficient \citep{Johansen2007,Carrera2015,Yang2017}. Here, certain conditions in protoplanetary discs (solid to gas ratio) might reduce the necessary particle size for concentration to occur \citep{Yang2017}.
First specific laboratory experiments accompanying simulations were initiated to verify these ideas in a sedimenting cloud of grains \citep{Lambrechts2016}. More recently, clouds of grains in a dilute gas levitated in a rotating setup for the first time actually showed a transition to sensitive collective behaviour \citep{Schneider2019}.
\\
\\
In any case, as a result the bouncing particles in protoplanetary discs are supposed to form even denser clumps of particles. If dense enough, they will slowly collapse under their own gravity \citep{Wahlberg2014}.
Being on the constructive side, this seems to be a promising way to build planetesimals as it avoids sticking problems and possibly overcomes the drift barrier \citep{Weidenschilling1977}. One might ask the question though if these weakly bound objects are really stable in protoplanetary discs. The interaction of solids and gas does not stop after planetesimal formation.

Once formed, planetesimals still move through the gas of the protoplanetary disc. Due to their large size, they do not couple to the gas, but move on Kepler orbits around the central star. As the gas is pressure supported a planetesimal likely experiences headwinds on the order of $50\,\rm m\,s^{-1}$ or more \citep{Weidenschilling1977,Johansen2014}.

\begin{figure}
	\begin{center}
		\includegraphics[width=.95\columnwidth]{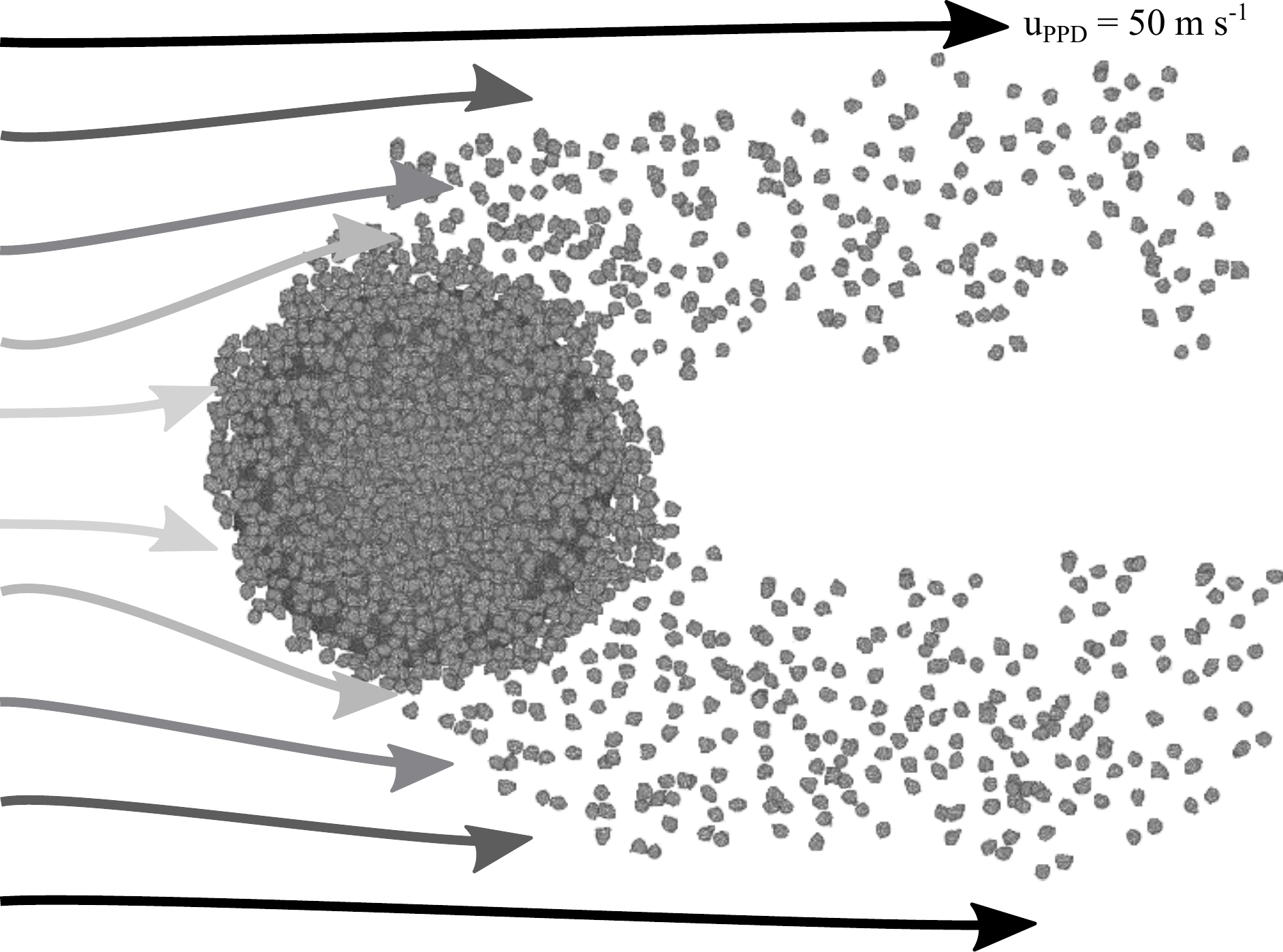}
	\end{center}
	\caption{\label{fig.idea} Planetesimals are destroyed under certain conditions of gas flow in protoplanetary discs. The relative velocity between the Kepler orbit of a planetesimal and the surrounding gas is $u_\mathrm{PPD}=50\,\rm m\,s^{-1}$. A planetesimal moving through the disk will affect the streaming gas. A planetesimal will be eroded if the gas density $\rho$ is high enough.}
\end{figure}

On planetary surfaces, wind is erosive. Compared to the Earth, the dynamic wind pressure in protoplanetary discs is lower due to lower gas densities. However, also the gravity of a kilometre sized object is low. In fact, the escape velocity is only on the order of $1\,\rm m\,s^{-1}$.
\citet{Paraskov2006} put forward the idea that dusty objects might be destroyed in few orbits if cohesion does not dominate. They supported their idea by ground based wind tunnel experiments.
Especially in view of weakly bound planetesimals, formed by gravitational instability, this idea gets even more important. The destructive case is sketched in fig. \ref{fig.idea}. Considering this, we started a new
series of experiments under low gravity to approach the problem of planetesimal stability in protoplanetary discs. Results and their implications are reported here.\\

We note that the parameter range of the experiments and the parameters of the application to planetesimal erosion are not yet overlapping. In this first set of experiments, we used as low gravity and as low pressure as was feasible for the existing wind tunnel $\rm (\sim 300\,Pa, 1.1\,m\,s^{-2})$.
The pressure is only at the upper end of the inner edge of models for protoplanetary discs, but most of the disc is orders of magnitude less dense. Also the gravity of a planetesimal is still much lower. So here, we test the validity of existing models and use extrapolations to estimate the stability of planetesimals. Refinements will require new experimental concepts but so far we consider our work as a first working approach.

The paper is structured as follows.
In section \ref{sec:pebblepileplanetesimals} we first give a short estimate on the stability of pebble pile planetesimals. Section \ref{sec:winderosion} introduces the models which currently exist to quantify the conditions for erosion on planetary surfaces by wind. In section \ref{sec:microgravityexperiments} we present our experimental results on wind erosion. The experiments probe and extend the validity range of the models to lower gravity and lower ambient pressure than studied before. In section \ref{sec:stabilityregions} we extrapolate the threshold conditions to planetesimals in protoplanetary discs and discuss the stability of planetesimals against wind erosion by protoplanetary winds. Section \ref{sec:conclusion} concludes the paper.

\section{Pebble Pile Planetesimals}
\label{sec:pebblepileplanetesimals}

In recent years it has become customary to call aggregates in the millimetre to decimetre size range pebbles. 
While it should be kept in mind that these objects are not stones but dust aggregates, some features taken for granted for pebbles might actually fit nicely. One of those is that pebbles do not stick together by cohesion easily. Without gravity a pebble pile would just dissolve at the lowest disturbance. This is also true for dust aggregates ("pebbles") at the bouncing barrier. They are compact and if the collision energy is not sufficient to fragment them, they essentially stay as they are and can easily be separated into a cloud of individual dust pebbles \citep{Brisset2013,Kothe2013}.
\citet{Wahlberg2017} simulated the collapse of a pebble cloud and found that small clouds forming planetesimals of up to $30\,$km size collapse into porous pebble piles. For those small collapsing pebble clouds the collision energies between the dust aggregates are not high enough to fragment them. Thus, we assume that small pebble pile planetesimals should have a low tensile strength and related to this a low shear strength. Only the inner part of larger bodies reaches a limit where dust pebbles are compressed into stronger dust layers by gravity.\\

Estimated tensile strengths are of the order of a few Pa. \citet{Skorov2012} and \citet{Blum2014} did experiments with dust granules confirming the low tensile strengths.
On even smaller scales of sub-millimetre aggregates composed of solid sub-millimetre grains the tensile strength was measured for the first time by \citet{Musiolik2017} to be between $10$ to $100\,$Pa. So even on smaller scales the tensile strength is rather low once the particle entities get larger than $100\, \mu$m. The other way around, if pebbles are much larger than centimetre size, the tensile strength might be well below $1\,$Pa. This is in agreement with the general treatment of particle systems consisting of larger grains as cohesionless granular medium.

This might be compared to the tensile strength generated by the self gravity of a planetesimal. The gravitational acceleration on the surface of a planetesimal with radius $R$ and a uniform density $\rho_\mathrm{Pla}$ can be calculated with
\begin{equation}
\label{eq:gravityPlanetesimal}
g=\frac{G M_\mathrm{Pla}}{R^2} = \frac{4}{3} \pi G \rho_\mathrm{Pla} R.
\end{equation} with the gravitational constant $G$. In low depth $z$ below the surface of a planetesimal of radius $R$ with a density of about $\rho_\mathrm{Pla} = 1000 \, \rm kg\,m^{-3}$ the pressure is 
\begin{equation}
P_z =\rho_\mathrm{Pla} g z =\frac{4}{3} \pi G \rho_\mathrm{Pla}^2 R z.
\end{equation}
A pressure of $1\,$Pa on a $10\,$km body is only reached $0.36\,$m below the surface. In the first mm surface layer the gravitational pressure is in the order of $10^{-3}\,$Pa. For a thin surface layer of millimetre to centimetre grains tensile strength due to cohesion dominates. For somewhat larger planetesimals both gravity and cohesion matter, eventually \citep{Greeley1980,Scheeres2010}. It has to be noted that the relevant pressures are very low. In conclusion  this implies that if the head wind can provide a drag related pressure of only $1\,$Pa, planetesimals can be eroded.\\

It is e.g. not clear how large cohesion under low gravity between dust pebbles really is. \citet{Musiolik2018} showed that wind erosion under microgravity proceeds at lower wind speed compared to ground based experiments, not only due to reduced gravity. But also cohesion was reduced strongly due to different packing of a forming surface under low gravity.

At low gas pressure the gas flow around the pebbles is no longer following the rules of continuum flow and it has not been studied yet how this influences wind erosion. This can actually only be accomplished under low gravity where the expected threshold velocities for wind erosion are lower than on Earth.

\section{Wind Erosion}
\label{sec:winderosion}

There are several models that predict the conditions for wind erosion. They are all going back to Bagnold's pioneering work \citep{Bagnold1941}. The idea behind his model is that lift occurs if the gas drag force on the particle is greater than the particle's gravitational force. The threshold friction velocity is then described by
\begin{equation}
\label{eq:bagnold}
u^{*}_\mathrm{B} = A \sqrt{\frac{\rho_\mathrm{p}-\rho}{\rho} g d} \stackrel{\rho_\mathrm{p} \gg \rho}{\approx} A \sqrt{\frac{\rho_\mathrm{p}}{\rho} g d}.
\end{equation}
$\rho_\mathrm{p}$ is the particle density, $d$ the particle diameter and $g$ the gravitational acceleration. The coefficient $A$ is called the dimensionless threshold friction velocity and depends on the inter-particle forces of the particles within the dust bed and the Reynolds number
\begin{equation}
\label{eq:ReynoldsNumber}
\mathrm{Re}^*=\frac{u^* d \rho}{\eta}
\end{equation}
of the flow with respect to the particles. $\eta$ is the dynamic viscosity of the gas. \citet{Iversen1982} extended the threshold friction velocity to 
\begin{equation}
\label{eq:iversen}
u^{*}_\mathrm{IW} = A_1 f(\mathrm{Re}^*)\sqrt{1 + \frac{K}{\rho_\mathrm{p} g d^n}}\sqrt{\frac{\rho_\mathrm{p}}{\rho} g d},
\end{equation}
with a constant $A_1$ and a function $f(\mathrm{Re}^*)$ that only depends on $\mathrm{Re}^*$. $K$ describes the inter-particle interaction. Apart from the dimension, $K$ is similar to the surface energy $\gamma$ of the particle material. The trend of $f(\mathrm{Re}^*)$ was determined experimentally \citep{Iversen1982}. For small Reynolds numbers $f(\mathrm{Re}^*)$ is nearly constant. \citet{Shao2000} derived a new expression for the threshold friction velocity
\begin{equation}
\label{eq:shao}
u^{*}_\mathrm{SL} = A_\mathrm{N}\sqrt{\frac{\rho_\mathrm{p}}{\rho} g d + \frac{\gamma}{\rho d}},
\end{equation}

which is a simplification of existing works on wind induced erosion. The authors argue with the weak $\mathrm{Re}^*$ dependence for particles with diameters between $30-1300\, \mu$m, so the dimensionless threshold friction velocity can be expressed as a constant $A=A_N$.

\section{Microgravity Experiments}
\label{sec:microgravityexperiments}

\subsection{Experimental Setup}

\begin{figure}
	\centerline{\includegraphics[width=\columnwidth]{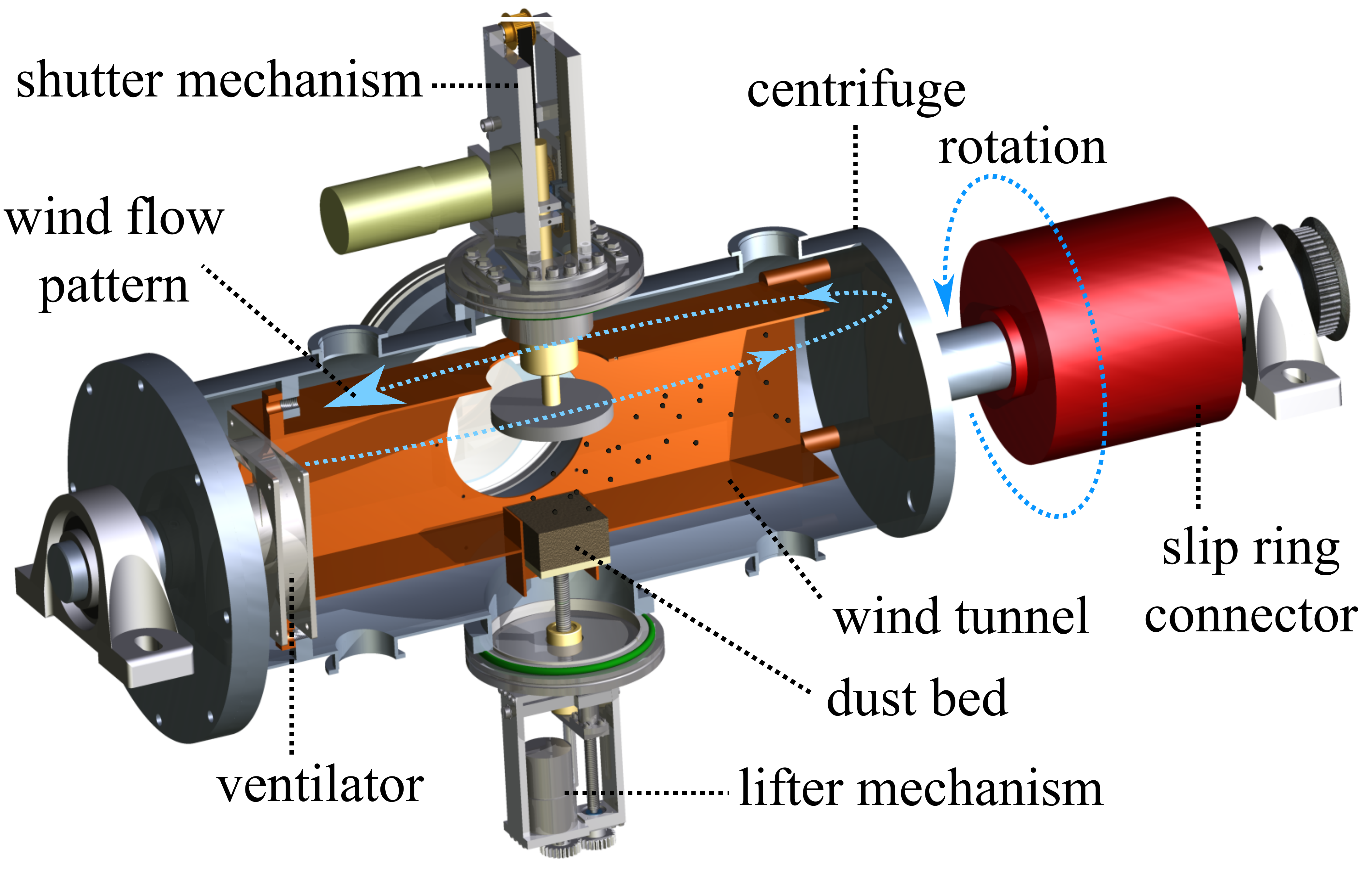}}
	\caption{A schematic of the experimental setup \citep{Musiolik2018}. The experiment combines a low pressure wind tunnel and a centrifuge. A ventilator inside the wind tunnel generates a gas flow with velocities of up to $15\,\rm m\,s^{-1}$.}
	\label{fig:setup}
\end{figure}

\begin{figure*}
	\centerline{\includegraphics[width=\textwidth]{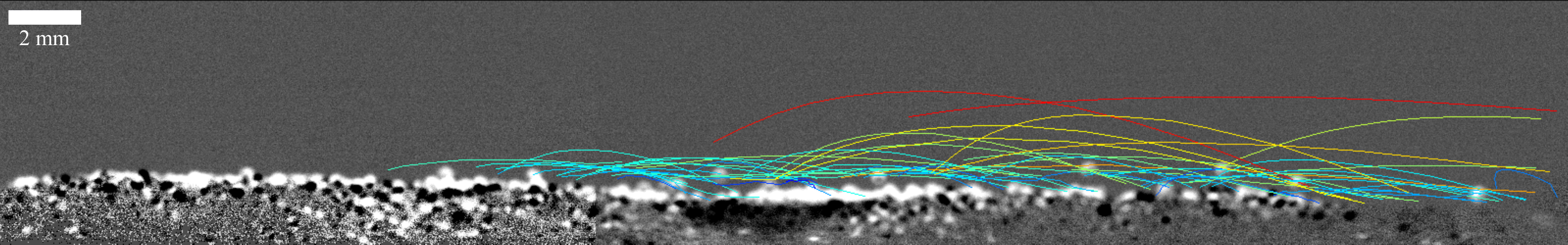}}
	\caption{Snapshot of the eroded bed at 0.19\,g. For a better visualization of the moving spheres (bright) a flat field division is applied. Additionally the trajectories of lifted spheres are overlayed. The colour indicates the maximum velocity of a particle within the track (blue$\rightarrow$red: slow$\rightarrow$fast). To exclude edge and wind shadow effects, the glass spheres located on the first centimetre (from the left) of the dust bed are not considered for the trajectory analysis. It takes about $1\,$s to erode the top layer of the dust bed.}
	\label{fig:rawImage}
\end{figure*}

\begin{figure*}
	\begin{minipage}{0.49\textwidth}
		\centerline{\includegraphics[width=\textwidth]{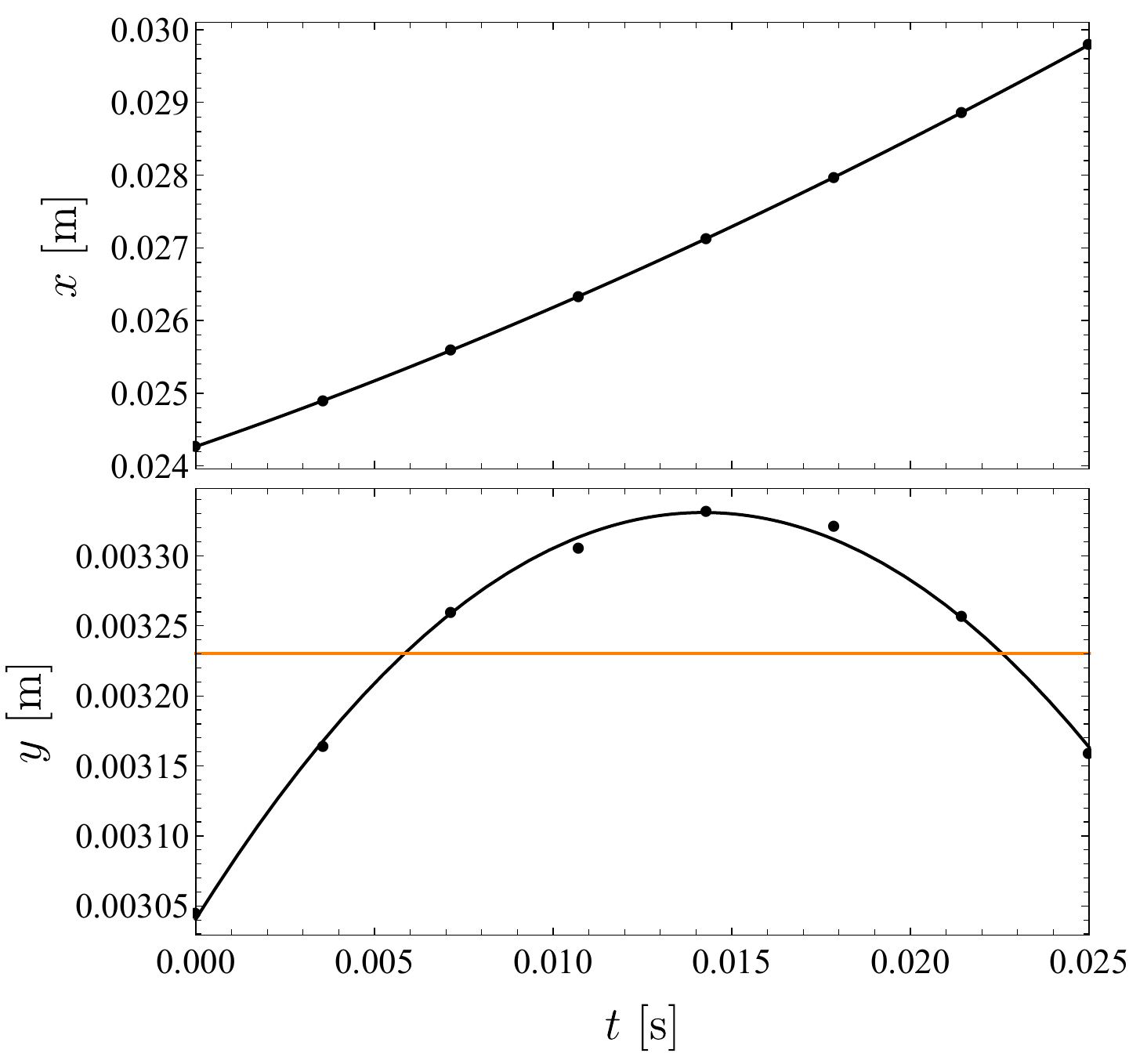}}
	\end{minipage}
	\begin{minipage}{0.49\textwidth}
		\centerline{\includegraphics[width=\textwidth]{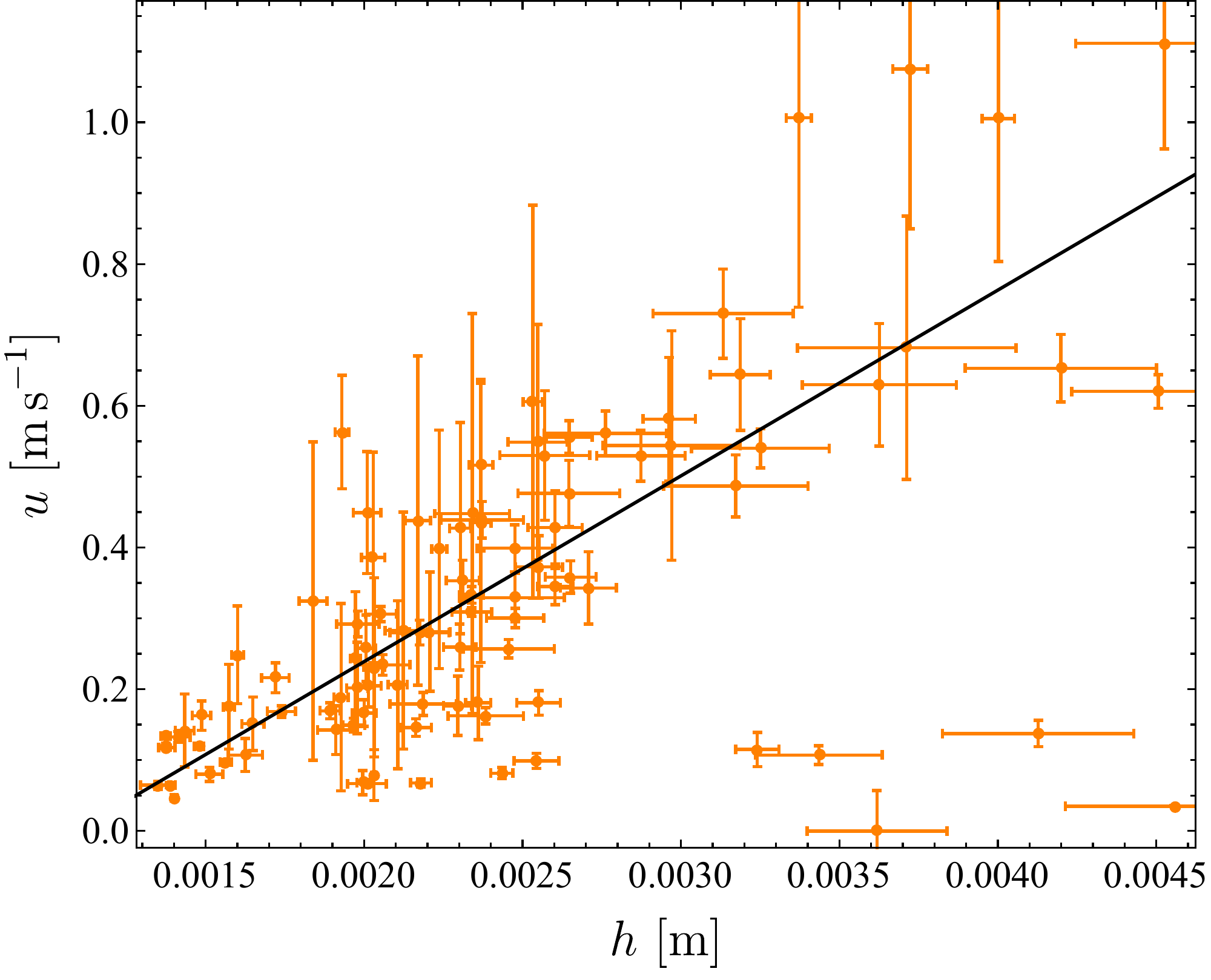}}
	\end{minipage}
	\caption{(Left) Example of the trajectory of a lifted bead at 0.19\,g: Eq. \ref{eq:x} is fitted to the $x$-component to estimate the gas velocity at a certain height. The corresponding height (see orange line) is determined by the mean value of the $y$-data within the fit range. Eq. \ref{eq:y} is fitted to the $y$-component. (Right) Wind profile of an experimental run: Close to the wall the wind velocity increases linearly with height. The illustrated error bars are originated from the fits.}
	\label{fig:trajectoryExample}
\end{figure*}

We modified the parabolic flight experiment used by \cite{Musiolik2018} that combines a low pressure wind tunnel and a centrifuge. A schematic of the experimental setup is shown in fig. \ref{fig:setup}. The wind tunnel is placed inside a vacuum chamber so it can be operated at various ambient pressures $p$ from $10$ to $10^5$\,Pa. The fan (max. frequency: $11.000\, \rm rpm$, max. air flow rate: $570\,\rm m^3 \, h^{-1}$) inside the wind tunnel generates a gas flow with velocities of up to $15\,\rm m\,s^{-1}$ in the centre of the wind tunnel. Both, the fan and the wind tunnel have a cross section of $10\,\mathrm{cm} \times 10\,\mathrm{cm}$. With the gas densities and maximum gas velocities studied in this work the Reynolds number of the wind tunnel is about $\mathrm{Re}_\mathrm{wt} \le 200$. Therefore, the wind flow inside the wind tunnel is laminar. The wind tunnel is smaller than the encapsulating vacuum chamber. Therefore, the gas streams through the wind tunnel and back again between the chamber walls at the outer side of the wind tunnel. This circulation is indicated in fig. \ref{fig:setup}. With this experiment we study the wind induced lift of dust and fine sand at different gravitational accelerations. The dust bed is placed at the bottom of the tunnel and is observed with a camera (see fig. \ref{fig:rawImage}) with a frequency of $280$\,Hz. The dust bed has a surface area of $4\,\mathrm{cm} \times 4\,\mathrm{cm}$. The chamber which contains the wind tunnel also acts as a centrifuge. During microgravity in a parabolic flight accelerations on the dust bed from 0.05 to 1\,g are generated with the centrifuge simulating different gravitational accelerations. To generate gravitational accelerations below $1\,$g, the experiment has to be carried out during a parabolic flight. Since we intend to simulate the wind erosion of a planetesimal surface, it is necessary to go down to gravitational accelerations as low as possible.\\

Our motivation for microgravity experiments in general is the dominant role of gravity over cohesion for sub-mm particles at 1g and therefore the unknown details of the cohesion dominated low gravity case. As mentioned, \citet{Musiolik2018} see a reduced cohesion for $\sim 100 \rm \,\mu m$ grains at Martian gravity compared to expectations due to different packings of settling grains. Therefore, gravity should be similar or rather less than the magnitude of the cohesion force $F_\mathrm{c}^\mathrm{JKR}=\frac{3}{2} \pi \gamma d$ \citep{Johnson1971}. For surface energies of about $\gamma = 10^{-4}\,\rm N\, m^{-1}$ (see results) gravity should be around $2\, \rm m\, s^{-2}$ at least, though the influence of gravity on cohesion and cohesion itself are not known \textit{a priori}. Gravity should therefore be as low as possible for these studies.\\

This experimental setup was used by \citet{Musiolik2018} under Martian conditions. The camera is mounted on the rotating chamber and is observing the dust bed through a window in perpendicular direction to the wind flow. At the beginning of each microgravity phase, the shutter opens and the dust bed is lifted up to expose the surface to the wind. The shutter as well as the lifter mechanism are located on the centrifuge. For a certain wind flow, gas pressure and centrifuge frequency, the dust bed is recorded for about $15\,\rm s$ with a camera frame rate of $280\,$Hz.

On the recent parabolic flight campaign we have determined the threshold wind velocity $u^*$ for spherical glass beads of diameter $d=425-450\,\mu$m for gravitational accelerations between 0.11 and 0.22\,g and ambient pressures between $300$ and $1200\,$Pa. The particle mass density is $\rho_\mathrm{p}=2460 \pm 30 \rm \, kg\,m^{-3}$. We chose this sample, because it is comparable with almost cohesionless pebbles on planetesimals. The particle size is similar to that of dust aggregates at the bouncing barrier. Assuming a random loose packing, the density of the dust bed is comparable to the mass density of a small planetesimal ($\rho_\mathrm{Pla} = 1000 \, \rm kg\,m^{-3}$). The probed gravitational accelerations and ambient gas pressures are the lowest which are technically possible with this experimental setup.\\
The determination of the threshold friction velocity $u^*$ was done by setting the gas flow high enough for lifting events to occur. Consistent with measurements given below, we consider a linear height dependence of the flow velocity $u(h)$ close to the wall, which is also present in viscous sublayers of turbulent flows \citep{Sternberg1962}. In detail we might rather have a Hagen-Poiseuille like parabolic flow but we cannot resolve this and, within the accuracy of the measurements, we consider the flow to be linear close to the wall. With a linear height dependence the threshold friction velocity is calculated as \citep{Schlichting2006}
\begin{equation}
\label{eq:thresholdVelocity}
u^*=\sqrt{\frac{\eta}{\rho}\frac{\partial u(h)}{\partial h}}.
\end{equation}

\subsection{Trajectory Analysis}

The height dependent gas flow velocity can be determined by the analysis of trajectories of lifted beads (particle tracking velocimetry). We used the \textit{Fiji} plug-in \textit{TrackMate} as tracking tool \citep{Schindelin2012,Tinevez2017}. In flow direction the motion is described by \citep{Wurm2001}
\begin{equation}
\label{eq:x}
x(t,h)=\left( u(h)-v_{0,x} \right) \tau \exp\left(-\frac{t}{\tau}\right)+u(h)t+x_0,
\end{equation}
where $\tau$ describes the constant coupling time and $x_0$, $v_{0,x}$ the
initial position and velocity of the sphere.\\
Fig. \ref{fig:rawImage} shows a snapshot of the eroded dust bed overlayed with the trajectories of the lifted beads. An example trajectory is shown in fig. \ref{fig:trajectoryExample}. Eq. \ref{eq:x} is fitted to the data. The corresponding height $h$ of the determined wind velocity $u(h)$ is the mean $y$-position of the fit. In order to determine the wind velocity of a certain height, the fit range is kept as narrow as possible (see fig. \ref{fig:trajectoryExample}, y-scale). In y-direction we fitted a parabolic equation which contains the simulated gravitational acceleration $g$ acting on the glass beads
\begin{equation}
\label{eq:y}
y(t)=-\frac{1}{2}g t^2+v_{0,y} t+y_0.
\end{equation}
Fig. \ref{fig:trajectoryExample} shows an example of the wind profile near the surface. The gas velocity $u(h)$ increases linearly with increasing height $h$. The resulting threshold friction velocity $u^*$ can be calculated with eq. \ref{eq:thresholdVelocity}.
\begin{figure}
	\centerline{\includegraphics[width=\columnwidth]{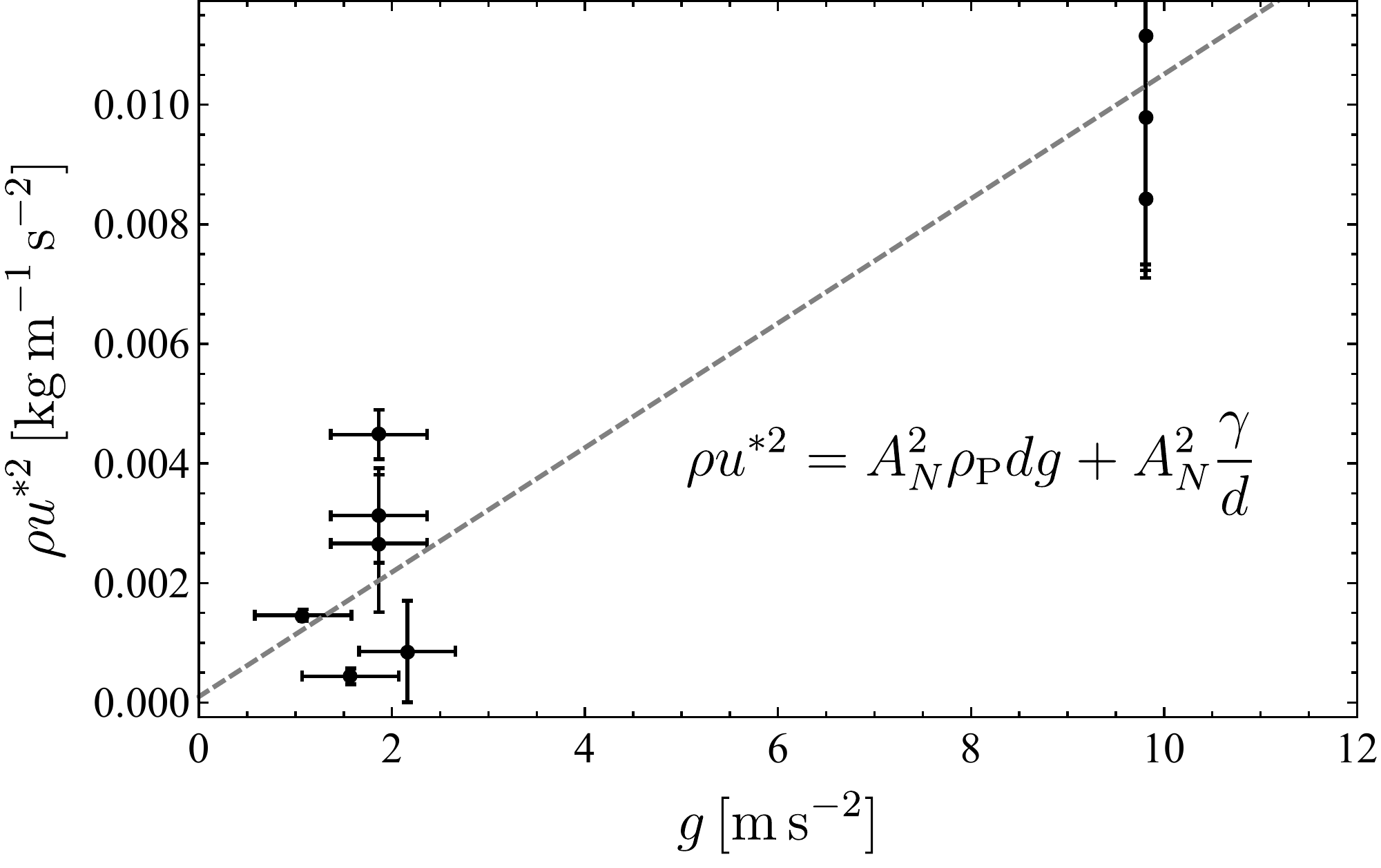}}
	\caption{$\rho u^{*2}$ in dependence of $g$ to determine the dimensionless threshold friction velocity $A_N$. Underlying the erosion model of \citet{Shao2000} the cohesion term is almost negligible for the spheres used in the experiment ($d=425-450\,\mu$m). The large scattering of the data is to be noted.}
	\label{fig:force}
\end{figure}

We note that due to the height dependence of the flow these are approximations, which we consider appropriate in the context of this study as the data are fitted well.

\subsection{Experimental Results}

In the recent parabolic flight campaign we have determined in total six threshold velocities for gravitational accelerations near $g\sim 2\,\rm m\,s^{-2}$ and ambient pressures $p$ between $300$ and $1200$\,Pa. As a comparison, three values were measured with the same experimental setup at Earth's gravity in a ground experiment. The results are summarized in tab. \ref{tab:results} and graphically shown in fig. \ref{fig:force}. Here $\rho u^{*2}$ is plotted against the gravitational acceleration $g$, thus the lift force and the gravitational force can be compared. We can put eq. \ref{eq:shao} into the form
\begin{equation}
\label{eq:shao2}
\rho u^{*2} = A_N^2 \left(\rho_\mathrm{p} g d + \frac{\gamma}{d} \right),
\end{equation}
where the left side describes the gas properties and is proportional to the lift force. The other side of the equation is the sum of gravity and cohesion which must be overcome for wind erosion to occur. If we fit our data with eq. \ref{eq:shao2}, we can determine the cohesion term ($y$-ordinate) and the dimensionless threshold friction velocity $A_N$ (gradient). We obtain $A_N=0.035\pm0.006$ and an almost negligible cohesion term $\gamma\leq3\times10^{-4} \,\rm N\,m^{-1}$. For spheres of the diameter $d=425-450\,\mu$m gravity is apparently the dominating force. The large scattering of the data is to be noted (see fig. \ref{fig:force}). This results from the fact that $u^*$ is experimentally determined in discrete steps and due to the g-jitter of the plane. Another reason might be the detailed shape of the dust bed, which cannot always be prepared in the exact same way.\\
We observe that it takes about $1\,$s ($\pm 0.3\,$s) to erode the top layer of the dust bed at the wind speeds close above the threshold friction velocities and thus the erosion rate near the erosion threshold is about  $0.6 \pm 0.2 \,\rm kg\,m^{-2}\,s^{-1}$. The erosion rate in dependence of the gas flow would require measurements at higher wind speeds than probed in this work.

\begin{table}
	\caption{Experimentally determined threshold friction velocities for the variation of gravitational acceleration $g$ and gas pressure $p$ or gas density $\rho$.}
	\label{tab:results}
		\begin{tabular}{cccc}
			\hline
			Gravitational & Gas & Gas & Threshold\\
			Acceleration & Pressure & Density & Friction Velocity\\
			$g~[\rm m\,s^{-2}]$ & $p~[\rm Pa]$ & $\rho~[\rm kg\,m^{-3}]$ & $u^*~[\rm m\,s^{-1}]$\\
			\hline
			$1.9 \pm 0.5$ & 300 & 0.0035 & $0.87 \pm 0.19$\\
			$1.9 \pm 0.5$ & 700 & 0.0082 & $0.74 \pm 0.03$\\
			$1.1 \pm 0.5$ & 710 & 0.0082 & $0.42 \pm 0.01$\\
			$1.9 \pm 0.5$ & 1180 & 0.0137 & $0.48 \pm 0.06$\\
			$2.2 \pm 0.5$ & 1120 & 0.0130 & $0.26 \pm 0.13$\\
			$1.6 \pm 0.5$ & 1010 & 0.0118 & $0.19 \pm 0.03$\\
			$9.8$ & 400 & 0.0046 & $1.55 \pm 0.27$\\
			$9.8$ & 720 & 0.0084 & $1.00 \pm 0.08$\\
			$9.8$ & 790 & 0.0092 & $1.03 \pm 0.13$\\
			\hline
		\end{tabular}
\end{table}

\section{Stability Regions in Protoplanetary Disks}
\label{sec:stabilityregions}

\begin{figure}
	\centerline{\includegraphics[width=\columnwidth]{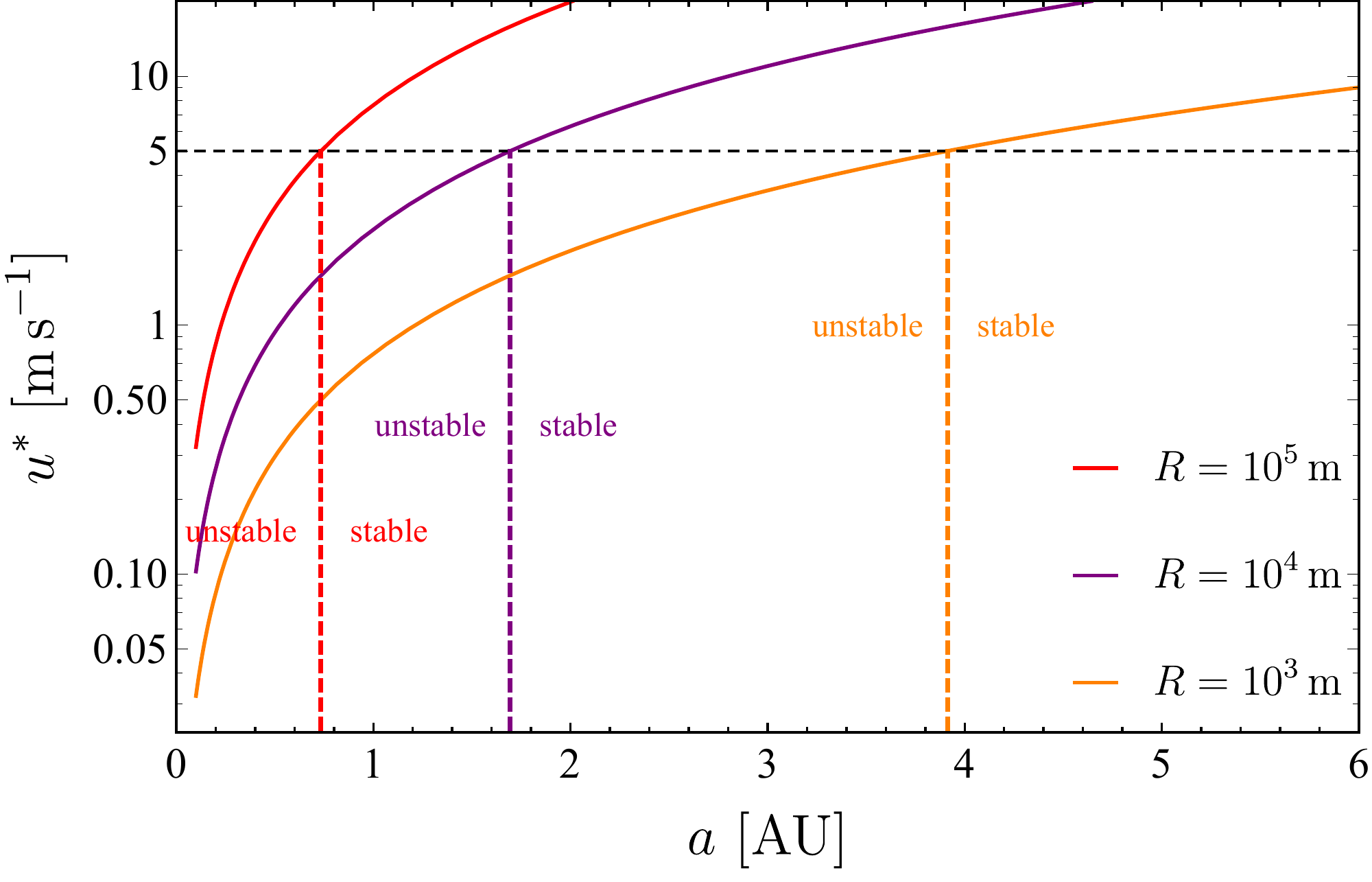}}
	\caption{$u^{*}$ in dependence of the radial distance to the star calculated for 3 different planetesimals in the minimum mass solar nebula with cohesionless pebbles ($\gamma=0$). The larger a planetesimal, the higher the threshold friction velocity for wind erosion.}
	\label{fig:uPPD}
\end{figure}

To apply the erosion model of \citet{Shao2000} on planetesimals, we need to know the gravitational acceleration on the surface of those objects and the local gas density in the protoplanetary disc. For the Minimum Mass Solar Nebula the gas density in dependence of the radial distance $a$ to the central star is described by \citep{Hayashi1981}
\begin{equation}
\label{eq:hayashi}
\rho=\rho_0 \left(\frac{a}{a_0}\right)^{-\frac{11}{4}},
\end{equation}
with $\rho_0= 1.4 \times 10^{-6}\,\rm kg\,m^{-3}$ and $a_0=1\,$AU. Fig. \ref{fig:uPPD} shows the needed threshold friction velocities in dependence of the radial distance to the star for planetesimals with different planetesimal radii and thus different gravitational accelerations.
Combining eq. \ref{eq:gravityPlanetesimal} and \ref{eq:shao2} with \ref{eq:hayashi} we can derive an expression for the stability threshold $a$ to the central star depending on the planetesimal radius
\begin{equation}
\label{eq:stabilityRegion}
a=a_0 \left[  \frac{A_N^2}{\rho_0 u^{*2}} \left( \frac{4}{3} \pi G \rho_\mathrm{p} \rho_\mathrm{Pla} R d +\frac{\gamma}{d}  \right) \right]^{-\frac{4}{11}}.
\end{equation}
Closer to the central star than the threshold semi-major axis $a$, a planetesimal of given size is unstable against wind erosion.

\begin{figure}
	\centerline{\includegraphics[width=\columnwidth]{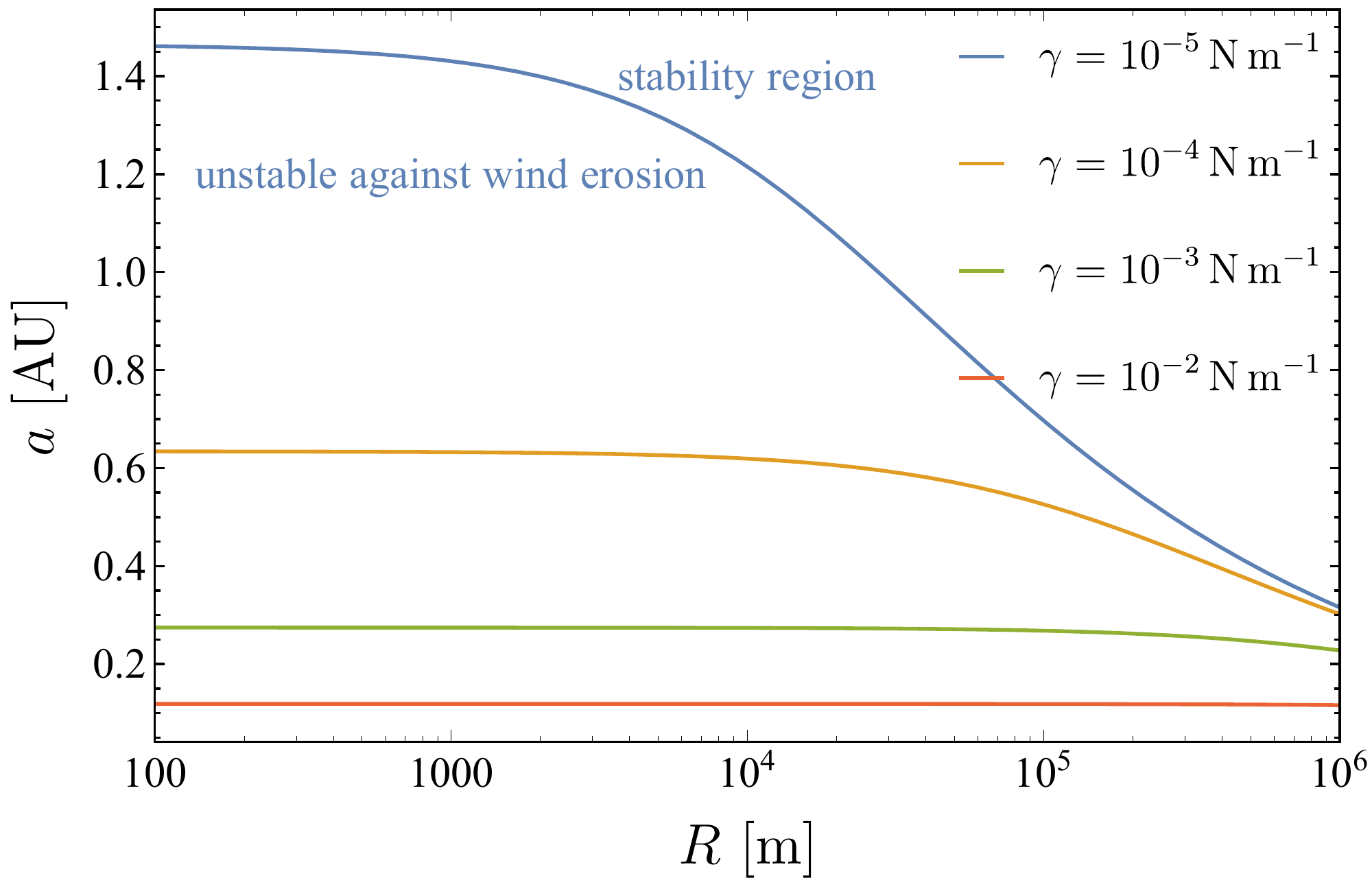}}
	\caption{Stability thresholds in the minimum mass solar nebula for a friction velocity of $u^* = 5\,\rm m\,s^{-1}$. Below the curves for given effective surface energy $\gamma$, planetesimals are not stable.}
	\label{fig:stability_region}
\end{figure}

The trend of this equation is shown in fig. \ref{fig:stability_region} for different surface energies $\gamma$. Below the curves planetesimals are unstable against wind erosion. We used $u^* = \alpha \cdot u_\mathrm{PPD}= 5\, \rm m\,s^{-1}$ here \citep{Greeley1980}. This value is a downscaling of the free flow gas velocity $u_\mathrm{PPD}=50\,\rm m\,s^{-1}$ relative to a planetesimal. This assumes the ratio between friction velocity and free flow velocity to be $\alpha=0.1$. This is uncertain at the low pressure of protoplanetary discs and currently unknown. It is conservative as it is taken from high pressure situations \citep{Greeley1980}. At lower ambient gas pressure slip flow might become important and shift the factor $\alpha$ to higher values. In that case the stability region of planetesimals would be smaller. The shift of eroding capability will certainly be of high importance on a local scale and to evaluate the potential of growing planetesimals at a certain location. However, the density dependence of the gas on radial distance to the star spans orders of magnitude. Also, the variations in gravity from small to large planetesimals span orders of magnitude. As we only outline the general eroding concept of the planetesimal in the protoplanetary disc here, the shift in flow regimes does not change the general picture, even if the friction velocity would be a factor of a few off. So at this point, we refrain from giving accurate predictions but consider the extrapolation as appropriate to highlight the concept which has not been proposed before.

\section{Conclusions}
\label{sec:conclusion}

Our results show that the erosion model by \citet{Shao2000} is applicable for the studied gas pressures $p$ and gravitational accelerations $g$ in the experiments. Based on this model the regions in protoplanetary discs which are unstable against wind erosion can be calculated. We derived an expression for the orbital distance $a$ which limits these regions (see eq. \ref{eq:stabilityRegion}). Closer to the central star than distance $a$ the gas pressure of the protoplanetary disc is high enough to erode a planetesimal with radius $R$ and pebble surface energy $\gamma$.

However, in our calculations we use a threshold friction velocity $u^*$ of $5\,\rm m\,s^{-1}$. This is roughly scaled down from the relative free flow velocity ($u_\mathrm{PPD}=50\,\rm m\,s^{-1}$) in protoplanetary discs. $u^*=u_\mathrm{PPD}$ would only be true if the planetesimal did not affect the streaming gas. In general, however, the planetesimal greatly influences the flow. The gas streams around the object and depending on the Reynolds number a wake develops. The gas velocity is smaller near the planetesimal's surface than the free flow velocity far away. For small $\mathrm{Re}$ the flow is laminar and stationary in time whereas for high $\mathrm{Re}$ the stream becomes turbulent. To understand the detailed gas streaming characteristics around a planetesimal, numerical simulations are needed. It also has to be noted that at the largest planetesimal size eroded grains no longer necessarily leave the planetesimal. A planetesimal will only be destroyed by the wind if the lifted particles do not fall back on the planetesimal but leave it. Therefore, the fraction of pebbles which leave the planetesimal by overcoming the planetesimal's gravitational potential barrier and the remaining fraction of redeposited pebbles would be of great interest. As the simplest estimation, we check if the lifted pebbles coupled to the gas can overcome the gravitational sedimentation velocity $v_\mathrm{sed}$. If, for simplicity, we treat the ambient gas as continuum fluid, the sedimentation velocity is described for small $\rm Re$ by the Stoke's law
\begin{equation}
\label{eq:sedimentation}
	v_\mathrm{sed}=\frac{1}{18} C g d^2 \frac{\rho_\mathrm{p}-\rho}{\mu}.
\end{equation}
It depends on the particle and gas density, the dynamic viscosity of the gas, the gravitational acceleration of the planetesimal, the pebble diameter and the Cunningham correction factor $C \approx 2.6$ to consider non-continuum effects at Knudsen number $\mathrm{Kn}\approx 1$ \citep{Davies1945}. If $v_\mathrm{sed}> u_\mathrm{PPD}=50\,\rm m\,s^{-1}$, the lifted pebble will fall back and will be redeposited somewhere on the planetesimal's surface. For a planetesimal with $10\,$km radius at $1\,$AU distance to the central star, this means that pebbles larger than approximately $2\,$cm will be redeposited on the planetesimal while smaller pebbles can leave the planetesimal. Indications of similar redeposit processes have been observed on the comet 67P/Churyumov-Gerasimenko \citep{Thomas2015,Jia2017}. In this context other processes like thermal fracturing might also be of importance to decrease the shear strength of icy planetesimals \citep{Kochan1989, Attree2018}\\

As another caveat it has to be mentioned that the parameters used in the experiments ($p$, $g$) are still in a regime where gravity dominates and the flow is hydrodynamic. For lower pressures, where the transition regime between the hydrodynamic and the molecular flow regime begins, and for lower gravity erosion is still unstudied. To understand the wind induced erosion of planetesimals, experiments at lower and more realistic gas pressures ($0.1-10\,$Pa) and low gravitational accelerations are needed. Therefore we are currently developing a new experimental setup working at these low pressures for gas velocities in excess of $50\,\rm m\,s^{-1}$ and accelerations down to $10^{-2}\,$g. We also plan to characterize the erosion rate at these realistic parameters with the new setup.

We note that pebble pile planetesimals are currently often considered to form  outside of the ice line for various reasons \citep{Drazkowska2017,Schoonenberg2017}. These authors argue, that water diffusion and its recondensation outside of the ice line significantly increase the density of water ice outside of the ice line, which can enhance the probability of planetesimal formation via streaming instability. \cite{Drazkowska2017}  also found that stickier and thus larger ice aggregates drift faster inwards than smaller aggregates composed of less stickier silicates which leads to the formation of so-called "traffic jams" at the ice line and thus to a density enhancement. This does not preclude the formation of planetesimals further inside. As pebble piles are built from the bouncing constituents, a formation in water ice dominated realms does not necessarily means that they are more stable, i.e. due to increased sticking of water. Also, recently \cite{Gundlach2018} found that at the low temperatures the tensile strength of water ice might not be that much different from silicates which translates to similar shear strengths for water ice and silicates. This also counts for other ice lines, e.g. the $\rm CO_2$ line as next major ice line as it also behaves essentially like silicates \citep{Musiolik2016}.

Staying outside the snowline, comets have been argued to have formed by gravitational collapse of pebble clouds \citep{Blum2017}. They would be safe in the outer system. Despite these yet unknown details, it is clear though, that pebble pile planetesimals with low shear strength are prone to destruction in the inner disk.

\section*{Acknowledgements}

This project is funded by DLR (Deutsches Zentrum f\"{u}r Luft- und Raumfahrt) space administration with funds from the BMWi (Bundesministerium f\"{u}r Wirtschaft und Energie) under grant 50 WM 1760. FJ is funded by DLR space administration with funds from the BMWi under grant 50 WM 1762. MK is funded by DFG (Deutsche Forschungsgemeinschaft) grant WU 321/14-1. NS is funded by DFG grant WU 321/16-1. We thank J. Kollmer for the constructive discussion. We also appreciate a thorough review by the anonymous referee.




\bibliographystyle{mnras}
\bibliography{demirci} 




%
%


\bsp	
\label{lastpage}
\end{document}